\documentclass[twoside]{article}
\usepackage[usenames,dvipsnames]{xcolor}
\usepackage{amsmath}
\usepackage{bm} 
\usepackage{xspace}
\usepackage{pifont}
\usepackage{appendix}
\usepackage{csquotes}
\usepackage{nicefrac}
\usepackage{listings}

\usepackage[ruled,vlined]{algorithm2e}
\SetKwRepeat{Do}{do}{while}
\SetKwComment{tcp}{$\rhd\;$}{}%

\usepackage{enumitem}
\usepackage{url}

\usepackage{amsfonts,amssymb,amsbsy,textcomp,marvosym,picins,amsmath,caption,threeparttable,amsthm,subfigure,float,lastpage,lscape}
\usepackage{eurosym,mathrsfs,fancyhdr,multicol,graphics,indentfirst,color,bm,upgreek,booktabs,graphicx,multirow,warpcol,epstopdf}
\usepackage[bookmarksnumbered=true,bookmarksopen=true,colorlinks=true,pdfborder=001,urlcolor=blue,linkcolor=black,anchorcolor=blue,citecolor=blue]{hyperref}
\usepackage[capitalise]{cleveref}

\newcommand{\parf}{\textsc{Parf}\xspace}
\newcommand{\eva}{\textsc{Frama-C/Eva}\xspace}
\newcommand{\framac}{\textsc{Frama-C}\xspace}
\newcommand{\framacsv}{\textsc{Frama-C-SV}\xspace}

\newcommand{\mopsa}{\textsc{Mopsa}\xspace}
\newcommand{\goblint}{\textsc{Goblint}\xspace}

\newcommand{\green}[1]{\textcolor{ForestGreen}{#1}}

\newcommand{\gray}[1]{\textcolor{Gray}{#1}}
\newcommand{\maroon}[1]{\textcolor{Maroon}{#1}}

\newcommand{\revision}[1]{#1}
\newcommand{\append}[1]{#1}
\newcommand{\best}[1]{{\textbf{#1}}}
\newcommand{\similar}[1]{{\underline{#1}}}
\newcommand{\dominant}[1]{{\textbf{#1}}}
\newcommand{\precision}{%
\texttt{-eva-precision} 0\xspace}
\newcommand{\default}{%
\textsc{Default}\xspace}
\newcommand{\expert}{%
\textsc{Expert}\xspace}
\newcommand{\official}{%
\textsc{Official}\xspace}
\newcommand{\parfopt}{%
\textsc{Parf\_Opt}\xspace}
\newcommand{\parfavg}{%
\textsc{Parf\_Avg}\xspace}

\newcommand{\selected}{%
\textsc{Selected}\xspace}
\newcommand{\excluded}{%
\textsc{Excluded}\xspace}

\newcommand{\bTRUE}{\ensuremath{\text{true}}}
\newcommand{\bFALSE}{\ensuremath{\text{false}}}
\newcommand{\meet}{\sqcap}
\newcommand{\join}{\sqcup}
\newcommand{\Pbase}{P_{\text{base}}}
\newcommand{\Pdelta}{P_{\text{delta}}}
\newcommand{\pbase}{p_{\text{base}}}
\newcommand{\pdelta}{p_{\text{delta}}}

\newcommand{\eeq}{~{}={}~}
\newcommand{\defeq}{{}\triangleq{}}
\newcommand{\ddefeq}{\ {}\defeq{}\ }
\newcommand{\qdefeq}{\quad{}\defeq{}\quad{}}
\newcommand{\ooplus}{~{}\oplus{}~}

\makeatletter
\renewcommand\paragraph{\@startsection{paragraph}{4}{\z@}%
    {3.25ex \@plus 1ex \@minus 0.2ex}%
    {-1em}%
    {\normalfont\normalsize\itshape}}
\makeatother

\renewcommand{\thefootnote}{\fnsymbol{footnote}}
\def\footnoterule{\kern 1mm \hrule width 10cm \kern 2mm}

\allowdisplaybreaks
\catcode`@=11
\def\title#1{\vspace{3mm}\begin{flushleft}\vglue-.1cm\Large\bf\boldmath\protect\baselineskip=18pt plus.2pt minus.1pt #1
\end{flushleft}\vspace{1mm} }
\def\author#1{\begin{flushleft}\normalsize #1\end{flushleft}\vspace*{-4pt} \vspace{3mm}}
\def\address#1#2{\begin{flushleft}\vglue-.35cm${}^{#1}$\small\it #2\vglue-.35cm\end{flushleft}\vspace{-2mm}\par}
\def\jz#1#2{{$^{\footnotesize\textcircled{\tiny #1}}$\footnotetext{$^{\footnotesize\textcircled{\tiny #1}}$#2}}}

\catcode`@=11
\def\section{\@startsection{section}{1}{\z@}%
 {-3ex \@plus -.3ex \@minus -.2ex}%
 {2.2ex \@plus.2ex}%
{\normalfont\normalsize\protect\baselineskip=14.5pt plus.2pt minus.2pt\bfseries}}
\def\subsection{\@startsection{subsection}{2}{\z@}%
 {-3ex\@plus -.2ex \@minus -.2ex}%
 {2ex \@plus.2ex}%
{\normalfont\normalsize\protect\baselineskip=12.5pt plus.2pt minus.2pt\bfseries}}
\def\subsubsection{\@startsection{subsubsection}{3}{\z@}%
 {-2.2ex\@plus -.21ex \@minus -.2ex}%
 {1.4ex \@plus.2ex}
{\normalfont\normalsize\protect\baselineskip=12pt plus.2pt minus.2pt\sl}}

\begin{document}
\thispagestyle{empty}
\vspace*{-13mm}
\noindent {\small Zhong-Yi Wang, Ming-Shuai Chen, Teng-Jie Lin {\it et al.}\ \textsc{Parf}: An Adaptive Abstraction-Strategy Tuner for Static Analysis.
JOURNAL OF COMPUTER SCIENCE AND TECHNOLOGY \ Vol.(No.): \thepage--\pageref{last-page}
\ Mon. Year. DOI 10.1007/s11390-015-0000-0}
\vspace*{2mm}

\title{P{\small ARF}: An Adaptive Abstraction-Strategy Tuner for Static Analysis}

\author{Zhong-Yi Wang${}^{1}$, Ming-Shuai Chen${}^{1,*}$, \emph{Senior Member, CCF}, Teng-Jie Lin${}^{1}$,\\ Lin-Yu Yang${}^{1}$, Jun-Hao Zhuo${}^{1}$, Qiu-Ye Wang${}^{2}$, Sheng-Chao Qin${}^{3}$,\\ Xiao Yi${}^{2}$, and Jian-Wei Yin${}^{1}$, \emph{Senior Member, CCF}}

\address{1}{College of Computer Science and Technology, Zhejiang University, Hangzhou 310012, China}
\address{2}{Fermat Labs, Huawei Inc., Dongguan 523000, China}
\address{3}{Guangzhou Institute of Technology, Xidian University, Xi'an 710071, China}

\vspace{3mm}

\noindent E-mail: wzygomboc@zju.edu.cn; m.chen@zju.edu.cn; tengjiecs@gmail.com; linyu.yang@zju.edu.cn;\\\indent\indent\ \ jhzhuo@zju.edu.cn; wangqiuye2@huawei.com; shengchao.qin@gmail.com; yi.xiao1@huawei.com;\\\indent\indent\ \ zjuyjw@zju.edu.cn\\[-1mm]

\noindent Received ; accepted .\\[1mm]

\let\thefootnote\relax\footnotetext{{}\\[-4mm]\indent\ Regular Paper}
\let\thefootnote\relax\footnotetext{{}\\[-4mm]\indent\ Special Section of ChinaSoft2024-Prototype}
\let\thefootnote\relax\footnotetext{{}\\[-4mm]\indent\ This work was supported by the Zhejiang Provincial Natural Science Foundation Major Program under Grant No.\ LD24F020013, the CCF-Huawei Populus Grove Fund under Grant No.\ CCF-HuaweiFM202301, the Fundamental Research Funds for the Central Universities of China under Grant No.\ 226-2024-00140, and the Zhejiang University Education Foundation's Qizhen Talent program.}
\let\thefootnote\relax\footnotetext{{}\\[-4mm]\indent\ $^*$Corresponding Author}
\let\thefootnote\relax\footnotetext{{}\\[-4mm]\indent\ \copyright Institute of Computing Technology, Chinese Academy of Sciences 2025}

\noindent {\small\bf Abstract} \quad  {\small
We launch {\parf} -- a toolkit for adaptively tuning abstraction strategies of static program analyzers in a fully automated manner. {\parf} models various types of external parameters (encoding abstraction strategies) as random variables subject to probability distributions over latticed parameter spaces. It incrementally refines the probability distributions based on accumulated intermediate results generated by repeatedly sampling and analyzing, thereby ultimately yielding a set of highly accurate abstraction strategies. {\parf} is implemented on top of {\eva} -- an off-the-shelf open-source static analyzer for C programs. {\parf} provides a web-based user interface facilitating the intuitive configuration of static analyzers and visualization of dynamic distribution refinement of the abstraction strategies. It further supports the identification of dominant parameters in {\eva} analysis. Benchmark experiments and a case study demonstrate the competitive performance of {\parf} for analyzing complex, large-scale real-world programs.
}

\vspace*{4mm}

\noindent{\small\bf Keywords} \quad {\small automatic parameter tuning, {\eva}, program verification, static analysis}

\vspace*{5mm}

\baselineskip=14.5pt plus.2pt minus.2pt
\parskip=0pt plus.2pt minus0.2pt
\begin{multicols}{2}

\section{Introduction}\label{sec:01-intro}

Static analysis is the process of analyzing a program without ever executing its source code. The goal of static analysis is to identify and help users eliminate potential {runtime errors} (RTEs) in the program, e.g., division by zero, overflow in integer arithmetic, and invalid memory accesses. Identifying an appropriate {abstraction strategy} -- for soundly approximating the concrete semantics -- is a crucial task to obtain a delicate trade-off between the accuracy and efficiency of static analysis: A finer abstraction strategy may yield fewer false alarms (i.e., approximation-caused alarms that do not induce RTEs) yet typically incurs less efficient analysis. State-of-the-art sound static analyzers, such as {\eva}~\cite{buhlerEvaFramaC27}, Astr{\'e}e~\cite{kastnerAbstractInterpretationIndustry2023}, {\goblint}~\cite{saanGoblintAutotuningThreadModular2023}, and {\mopsa}~\cite{tacas24:mopsa}, 
integrate abstraction strategies encoded by various external parameters, thereby enabling analysts to balance accuracy and efficiency by tuning these parameters.

Albeit with the extensive theoretical study of sound static analysis~\cite{cousotAbstractInterpretationUnified1977,venetGaugeDomainScalable2012}, the picture is much less clear on its {parameterization} front~\cite{blanchetStaticAnalyzerLarge2003}: it is challenging to find a set of high-precision parameters to achieve low false-positive rates within a given time budget. The main reasons are two-fold: 1) Off-the-shelf static analyzers often provide a wide range of parameters subject to a huge and possibly infinite joint parameter space. For instance, the parameter setting in Table 1 consists of 13 external parameters that are highly relevant to the accuracy and efficiency of {\eva}, among which eight integer parameters have infinite value spaces. 2) The process of seeking highly accurate results typically requires multiple trials of parameter setting and analysis, which generates a large amount of intermediate information such as RTE alarms and analysis time. Nevertheless, few static analyzers provide a fully automated approach to guiding the refinement of abstraction strategies based on such information. Therefore, the use of sound static analysis tools still relies heavily on expert knowledge and experience.
\tabcolsep 4pt
\renewcommand\arraystretch{1.3}
\noindent
\begin{center}
\vspace{-1mm}
{\footnotesize{\bf Table 1.} Parameter Settings in {\eva}\label{tab:parameter-setting}}\\
\vspace{2mm}
\footnotesize{
\begin{tabular*}{\linewidth}{lcc}\hline\hline\hline
Parameter & Type & Value space\\\hline
\verb|min-loop-unroll| & Integer & $\mathbb{N}$ \\
        \verb|auto-loop-unroll| & Integer & $\mathbb{N}$ \\
        \verb|widening-delay| & Integer & $\mathbb{N}$ \\
        \verb|partition-history| & Integer & $\mathbb{N}$ \\
        \verb|slevel| & Integer & $\mathbb{N}$ \\
        \verb|ilevel| & Integer & $\mathbb{N}$ \\
        \verb|plevel| & Integer & $\mathbb{N}$ \\
        \verb|subdivide-non-linear| & Integer & $\mathbb{N}$ \\
        \verb|split-return| & String & $\{\text{``''}, \text{``auto''}\}$ \\
        \verb|remove-redundant-alarms| & Boolean & $\{\bFALSE, \bTRUE\}$ \\
        \verb|octagon-through-calls| & Boolean & $\{\bFALSE, \bTRUE\}$ \\
        \verb|equality-through-calls| & String & $\{\text{``none''}, \text{``formals''}\}$ \\
        \verb|domains| & Set-of-Strings & $\{\bFALSE, \bTRUE\}^5$ \\
\hline\hline\hline
\end{tabular*}
\\\vspace{2mm}
}
\end{center}

Some advanced static analyzers attempt to address the above challenges using various methods. {\eva}~\cite{buhlerEvaFramaC27} provides the meta option \verb|-eva-precision|, which packs a predefined group of valuations to the parameters listed in Table 1, thus enabling a quick setup of the analysis. K{\"a}stner \textit{et al.}~\cite{kaestnerAutomaticSoundStatic2023} summarized the four most important abstraction strategies in Astr{\'e}e and recommended prioritizing the accuracy of related abstract domains, which amounts to narrowing down the parameter space. However, both {\eva} and Astr{\'e}e currently do not support automatic parameter generation. {\goblint}~\cite{saanGoblintAutotuningThreadModular2023} implements a simple, heuristic autotuning method based on syntactical criteria, which can automatically activate or deactivate abstraction techniques before analysis. However, this method only generates an initial analysis configuration once and does not dynamically adapt to refine the parameter configuration. See \cref{sec:06-related-work} for detailed related work.

Following this line of research, we have presented {\parf}~\cite{ase24-parf}, an adaptive and fully automated parameter refining framework for sound static analyzers. {\parf} models various types of parameters as random variables subject to probability distributions over latticed parameter spaces. Within a given time budget, {\parf} identifies a set of highly accurate abstraction strategies by incrementally refining the probability distributions based on accumulated intermediate results generated via repeatedly sampling and analyzing.
Preliminary experiments have demonstrated that {\parf} outperforms state-of-the-art parameter-tuning mechanisms by discovering abstraction strategies leading to more accurate analysis, particularly for programs of a large scale. 

\paragraph*{Contributions.}
This article presents the {\parf} artifact, whose theoretical underpinnings have been established in~\cite{ase24-parf}. We focus on the design, implementation, and application of the {\parf} toolkit and make -- in position to~\cite{ase24-parf} -- the following new contributions:

1) We present design principles underneath the novel abstraction-strategy tuning architecture {\parf} for establishing provable incrementality (monotonic knowledge retention) and adaptivity (resource-aware exploration) to achieve accuracy-efficiency tradeoffs in static analysis parameterization.

2) We develop a web-based user interface (UI) for {\parf} which facilitates the intuitive configuration of static analysis and visualizes the dynamic distribution refinement of abstraction strategies.

3) We show via a post-hoc analysis that {\parf} supports the identification of the most influential parameters dominating the accuracy-efficiency trade-off.

4) We demonstrate through a case study how {\parf} can help eliminate false alarms and, in some cases, certify the absence of RTEs.


\section{Problem and Methodology}\label{sec:02-method}

This section revisits the problem of abstraction-strategy tuning and outlines the general idea behind our {\parf} framework. More technical details are in~\cite{ase24-parf}.

In abstraction-strategy tuning, a {static analyzer} is modeled as a function $\textit{Analyze}\colon (\textit{prog}, p) \mapsto A_p$, which receives a target program $\textit{prog}$ and a parameter setting $p$ (encoding an abstraction strategy of the analyzer) and returns a set $A_p$ of RTE alarms emitted under $p$~\cite{ase24-parf}. We assume, as is the case in most state-of-the-art static analyzers~\cite{oopsla24-selecting-abstraction}, that the analyzer exhibits {monotonicity} over parameters, i.e., an abstraction strategy of {higher} precision (in an ordered joint parameter space) induces {fewer} alarms and thereby {more accurate} analysis.

The problem of abstraction-strategy tuning reads as follows.
Given a target program $\textit{prog}$, a time budget $T \in \mathbb{R}_{> 0}$, a static analyzer $\textit{Analyze}$, and the joint space of parameter settings $S$ of $\textit{Analyze}$, find a parameter setting $p \in S$ such that $\textit{Analyze}(\textit{prog}, p)$ returns as few alarms as possible within $T$~\cite{ase24-parf}.

Our {\parf} framework~\cite{ase24-parf} addresses the problem as follows. It models external parameters of the static analyzer as {random variables} subject to probability distributions over parameter spaces equipped with complete lattice structures. It {incrementally} refines the probability distributions based on accumulated intermediate results generated by repeatedly sampling and analyzing, thereby ultimately yielding a set of highly accurate parameter settings within a given time budget. More concretely, \parf adopts a multi-round iterative mechanism. In each iteration, \parf 1) repeatedly {samples} parameter settings based on the initial or refined probability distribution of parameters, then 2) uses these parameter settings as inputs to the static analyzer to {analyze} the program, and finally 3) utilizes the analysis results to {refine} the probability distribution of parameters. \parf continues this process until the prescribed time budget is exhausted, upon which it returns the analysis results of the final round together with the final probability distribution of parameters.


\revision{The core technical challenge lies in designing the {representation of probability distributions} over latticed parameter spaces and the {iterative refinement mechanism} that jointly enforce 1) incrementality: rewardless analyses (i.e., no new false alarms are eliminated) with low-precision parameters do not occur; and  2) adaptivity: analysis failures can be avoided while enabling the effective search of high-precision parameters. Specifically, we model each parameter as a combination of dual random variables ($\Pbase$ and $\Pdelta$) with type-specific initialization. Then we design $\Pbase$ and $\Pdelta$ stratified refinement strategies, respectively, which guarantees: 1) incremental $\Pbase$ expectation to preserve the accumulated knowledge during the iterative procedure, and 2) adaptive $\Pdelta$ expectation scaling to balance tradeoffs of exploring the uncharted parameter space and high-precision analysis resource costs. Related details are illustrated in \cref{sec:03-architecture}.}

Regarding implementation, we are primarily concerned with the open-source static analyzer {\eva}~\cite{buhlerEvaFramaC27} for C programs. However, since \parf treats the underlying analyzer as a black-box function, it can be integrated with any static analyzer exhibiting monotonicity (e.g., {\mopsa}~\cite{tacas24:mopsa} as shown in~\cite{ase24-parf}).
\begin{figure*}[t]
\centering
\includegraphics[width=\linewidth]{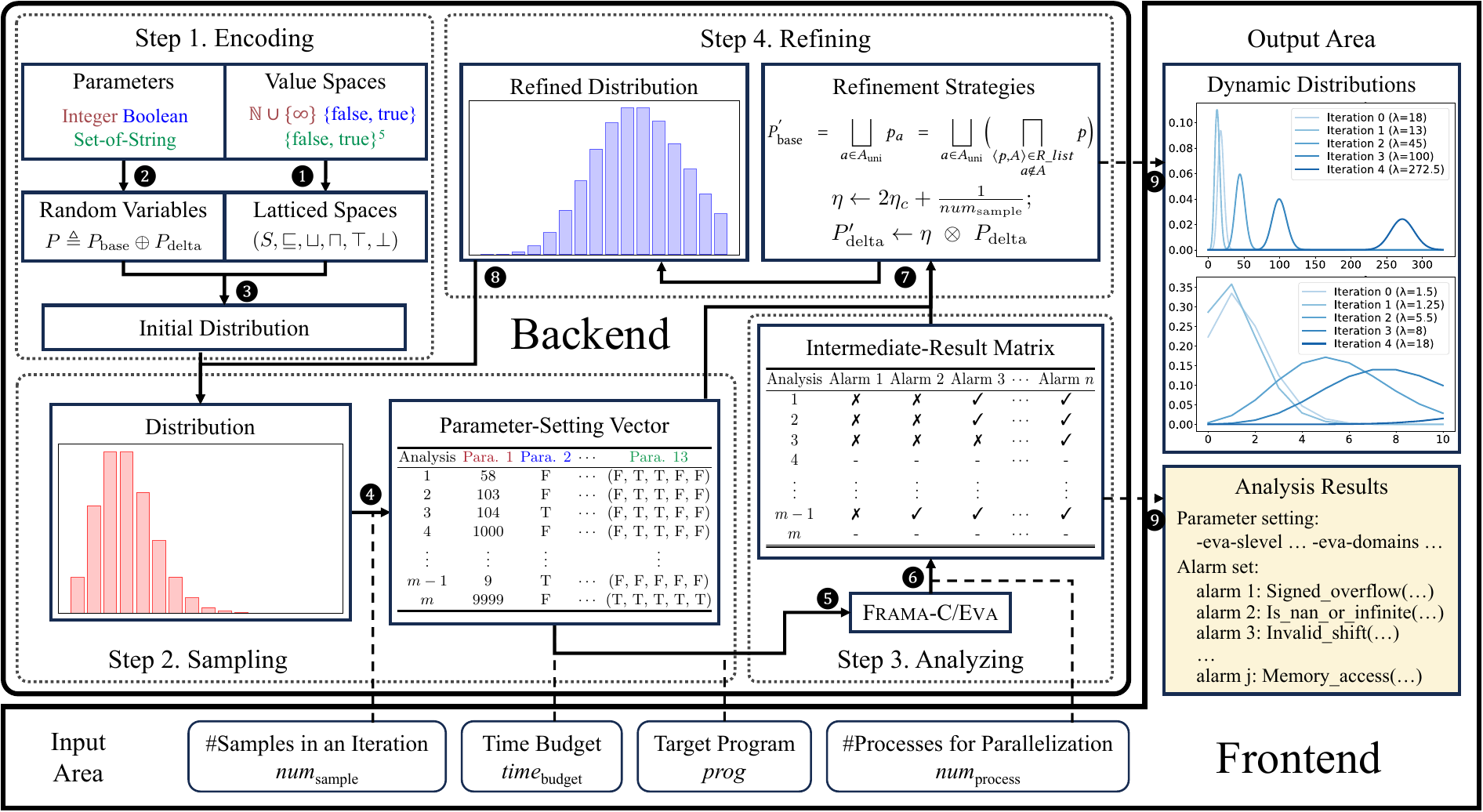}
\caption{Architecture of the {\parf} artifact.}\label{fig:01-overview}
\end{figure*}

\section{The P{\footnotesize ARF} Architecture}\label{sec:03-architecture}

This section elaborates on {\parf} artifact that implements the aforementioned techniques. As depicted in \cref{fig:01-overview}, the artifact is composed of two components: the backend tuning algorithm and the frontend web UI. The former comprises about 1,500 lines of OCaml code
~and the latter is built using Next.js and TypeScript.

The workflow of backend tuning algorithm comprises four main steps.

1) {Value-Space Encoding} (Subsection~\ref{subsubsec:01-encode}). {\parf} encodes the value spaces of parameters as sample spaces with complete lattice structures (\ding{182}). Meanwhile, it models external parameters of {\eva} as random variables subject to probability distributions over those latticed spaces (\ding{183}). This step aims to initialize the parameter distribution (\ding{184}), which serves as the basis for subsequent sample-analyze-refine iterations.

2) {Parameter Sampling} (Subsection~\ref{subsubsec:02-sample}). {\parf} repeatedly samples (\ding{185}) parameter settings as per either the initial distribution (from step 1) or the refined distribution (from step 4). The number of samples is determined by a user-defined hyper-parameter $num_\text{sample}$.

3) {Program Analyzing} (Subsection~\ref{subsubsec:03-analyze}). Using the parameter settings generated in step 2, {\parf} performs static analysis (\ding{186}) on the target program $\textit{prog}$ via {\eva} within the given time budget $time_\text{budget}$. The artifact supports parallelization and thus allows multiple analyses to be conducted simultaneously (\ding{187}). Once the time budget for this step is exhausted, {\parf} collects the intermediate results (e.g., the termination conditions and reported alarms) from each analysis and proceeds to the next step.

4) {Distribution Refining} (Subsection~\ref{subsubsec:04-refine}). {\parf} utilizes the intermediate results to refine the probability distribution (\ding{188}). It then returns to step 2, using the updated distribution as input (\ding{189}).

The frontend web-based UI enables intuitive and flexible interaction between users and the backend for, e.g., uploading target programs and configuring hyper-parameters \revision{($num_\text{sample}$, $time_\text{budget}$, and $num_\text{process}$)}. Moreover, the UI visualizes the dynamic evolution of parameter distributions during the analysis and displays the final analysis results along with the corresponding abstraction strategy (\ding{190}). Videos on accessing and using the UI are available online\jz{1}{\url{https://doi.org/10.5281/zenodo.13934703}, May 2025.}. Below, we explain each function module of the artifact in detail.

\setcounter{table}{2}
\tabcolsep 5pt
\renewcommand\arraystretch{1.3}
\begin{table*}[t]
\centering
\caption{\label{tab:03-random-variables-distributions} Distributions of $P$, $\Pbase$, and $\Pdelta$}\vspace{-2mm}
{\footnotesize
\begin{tabular*}{\linewidth}{ccccc}\hline\hline\hline
Type & $S$ & Distribution of $\Pbase$ & Distribution of $\Pdelta$ & $P \,=\, \Pbase\oplus\Pdelta$ \\\hline
Integer & $\mathbb{N}\cup\{\infty\}$ & $\Pr[\Pbase=a]=1$ & $\text{Poisson}(\lambda)$ & $a\,+\,\Pdelta$ \\
Boolean & $\{\bFALSE, \bTRUE\}$ & $\Pr[\Pbase=b]=1$ & $\text{Bernoulli}(q)$ & $b\,\lor\,\Pdelta$ \\
Set-of-Strings & $\{\bFALSE, \bTRUE\}^c$ & $\Pr[\Pbase=(b_1,\ldots,b_c)]=1$ & $\text{Bernoulli}(q_1)\times\ldots\times\text{Bernoulli}(q_c)$ & $(b_1\lor\Pdelta^1)\times\ldots\times(b_c\lor\Pdelta^c)$ \\\hline\hline\hline
\end{tabular*}
\\\vspace{1mm}\parbox{17.5cm}{Note: $\Pbase$ follows a Dirac distribution where $a$ is an integer sample and $b,b_1,\ldots,b_c$ are Boolean values. $\Pdelta$ adopts one of the following distributions, depending on the parameter type: Poisson distribution, Bernoulli distribution, or $c$-dimensional independent joint Bernoulli distribution (in this case, $\Pdelta$ can be expressed as $(\Pdelta^1, \ldots, \Pdelta^c)$). The binary operator $\oplus$ also varies based on the parameter type: it corresponds to addition ($+$) and logical disjunction ($\lor$) for an integer or Boolean parameter, respectively; for a set-of-strings parameter with cardinality $c$, $\oplus$ is defined as the point-wise lifting of $\lor$ to a $c$-dimensional random vector.}
}
\end{table*}
\baselineskip=18pt plus.2pt minus.2pt
\parskip=0pt plus.2pt minus0.2pt


\subsection{Value-Space Encoding}\label{subsubsec:01-encode}

Table 1 lists 13 parameters encoding \eva's abstraction and analysis strategies, categorized into four types with value spaces defined by their type: 1) integer parameters range over $\mathbb{N} $; 2) Boolean parameters have a value space of $\{\bFALSE, \bTRUE\}$; 3) string parameters have a value space defined as a set of strings; and 4) \verb|domains|, a unique set-of-strings parameter takes values from a power set of five abstract domains, namely, \{``cvalues'', ``octagon'', ``equality'', ``gauges'', ``symbolic-locations''\}. A value of \verb|domains| can be represented as a quintuple consisting of $\bTRUE$ or $\bFALSE$, indicating whether the corresponding domain is enabled. For example, the quintuple $(\bFALSE, \bFALSE, \bTRUE, \bTRUE, \bFALSE)$ corresponds to \{``equality'', ``gauges''\}. Thus, the value space of \verb|domains| is $\{\bFALSE, \bTRUE\}^5$.

\tabcolsep 6pt
\renewcommand\arraystretch{1.3}
\noindent
\begin{center}
\vspace{-1mm}
{\label{tab:02-latticed-sample-space}\footnotesize{\bf Table 2.} Latticed Sample Spaces of Different Parameter Types}\\
\vspace{2mm}
\footnotesize{
\begin{tabular*}{\linewidth}{cccccc}\hline\hline\hline
Type & $a\sqsubseteq b$ & $a\join b$ & $a\meet b$ & $\top$ & $\bot$ \\\hline
    Integer & $a\leq b$ & $\text{max}(a,b)$ & $\text{min}(a,b)$ & $\infty$ & $0$ \\
    Boolean & $a\Rightarrow b$ & $a\lor b$ & $a\land b$ & $\bTRUE$ & $\bFALSE$ \\
    Set-of-Strings & $a\subseteq b$ & $a\cup b$ & $a\cap b$ & $U$ & $\emptyset$ \\
\hline\hline\hline
\end{tabular*}
\\\vspace{1mm}\parbox{8.3cm}{Note: The elements $a$ and $b$ in each row are of their respective type, e.g., $a=2, b=5$ for integer parameter, $a=\bFALSE, b=\bTRUE$ for Boolean parameter, and $a=\{\text{``equality''}\}, b=\{\text{``equality''}, \text{``gauges''}\}$ for set-of-strings parameter.}
}
\end{center}

{\parf} encodes the value spaces of parameters into {latticed sample spaces}, represented as $(S, \sqsubseteq, \join, \meet, \top, \bot)$, where $S$ denotes the value space of a parameter, $\sqsubseteq$ is the {partial order} over $S$, $\join$ denotes the \emph{join} (aka the least upper bound) operator, $\meet$ denotes the \emph{meet} (aka the greatest lower bound) operator, $\top$ and $\bot$ stand for the {greatest} and {least} element in $S$, respectively. Table 2 instantiates these symbols for each parameter type. Note that the two string-typed parameters of {\eva} (cf.~Table 1) have only two possible values corresponding to two abstraction strategies with different precision levels, thus allowing us to treat them as Boolean-typed parameters. The lattice structure of the sample spaces serves as the basis of the distribution refinement mechanism described in Subsection~\ref{subsubsec:04-refine}.

{\parf} models each parameter as a composite random variable $P$ in the novel form of
\begin{align}\label{eq:composite-dist}
    P \qdefeq \Pbase \ooplus \Pdelta~,\tag{$1$}
\end{align}%
where $\Pbase$ is a {base} random variable for {retaining} the accumulated knowledge during the iterative analysis whilst $\Pdelta$ is a {delta} random variable for {exploring} the parameter space; they share the same sample space and range with $P$.
$\Pbase$ follows Dirac distribution, i.e., $\Pr[\Pbase = p] = 1$ for some sample $p \in S$; $\Pdelta$ adopts different types of distributions as per the parameter type: we use Bernoulli distributions for Boolean-typed parameters and Poisson distributions for integer-typed parameters (since the latter naturally encodes infinite-support discrete distributions over $\mathbb{N}$). The construction of $P$ by combining $\Pbase$ and $\Pdelta$ via the operator $\oplus$ is given in \cref{tab:03-random-variables-distributions}.

\paragraph*{\textit{Remark}.} Determining appropriate distributions for $\Pdelta$ presents a technical challenge. Boolean parameters naturally match Bernoulli distributions due to their binary support set. For integer parameters, we utilize Poisson distributions for two key reasons: 1) the Poisson distribution offers an infinite support set that aligns well with the nature of integers, and 2) the Poisson distribution is characterized by a unique parameter $\lambda$.

\subsection{Parameter Sampling}\label{subsubsec:02-sample}
{\parf} repeatedly samples values for each parameter represented as a random variable $P$ following the composite distribution as in~(\ref{eq:composite-dist}). For instance, the initial distributions employed by the artifact are collected in \append{Table 8 of Appendix~\cref{appx-A:initial-prob-dist}}.


To generate a sample point $p$ for parameter $P$, {\parf} first draws samples $\pbase$ and $\pdelta$ independently from the distributions of $\Pbase$ and $\Pdelta$, respectively, and then applies the binary operation $\oplus$ to construct the sampled value for $P$, i.e., $p = \pbase\oplus\pdelta$ (see \cref{tab:03-random-variables-distributions}). Subsequently, {\parf} aggregates the sampled values of all parameters into a complete analysis configuration. The total number of generated configurations in a sample-analyze-refine iteration is controlled by the user-defined hyper-parameter $num_\text{sample}$. All the configurations are maintained in an internal list structure.

\subsection{Program Analyzing}\label{subsubsec:03-analyze}
In this step, {\parf} performs static analysis on the target program $\textit{prog}$ leveraging {\eva}. The analyses pertaining to the $num_\text{sample}$ parameter settings obtained in the previous step are mutually independent and thus can be parallelized. However, {\eva} per se does not support the execution of parallel tasks. Hence, we implement this functionality using the OCaml module \texttt{Parmap}\jz{2}{\url{https://opam.ocaml.org/packages/parmap/}, May 2025}. The degree of parallelization, i.e., the number of processes, is determined by the user-defined hyper-parameter $num_\text{process}$.

Some analyses may fail to terminate within the given time limit, which is constrained by the total time budget (controlled by a hyper-parameter $time_\text{budget}$) for all the sample-analyze-refine rounds. For each analysis, {\parf} records whether it terminates and, if yes, the so-reported alarms. These intermediate results are then utilized to refine the distribution of $P$.

\subsection{Distribution Refining}\label{subsubsec:04-refine}

\tabcolsep 5pt
\renewcommand\arraystretch{1.3}
\noindent
\begin{center}
\vspace{-1mm}
{\footnotesize{\bf Table 4.} Example of Refining $\Pbase$ for \verb|slevel|}\\
\vspace{2mm}
\footnotesize{
\begin{tabular*}{\linewidth}{cccccc}\hline\hline\hline
        Analysis & Value & Alarm 1 & Alarm 2 & Alarm 3 & Alarm 4\\\hline
        1 & 58 & \ding{55} & \ding{55} & \ding{51} & \ding{51} \\
        2 & 103 & \ding{55} & \ding{55} & \ding{51} & \ding{51} \\
        3 & 104 & \ding{55} & \ding{55} & \ding{55} & \ding{51} \\
        4 & 1000 & \ding{55} & \ding{55} & \ding{55} & \ding{51} \\
        5 & 9 & \ding{55} & \ding{51} & \ding{51} & \ding{51} \\
        6 & 9999 & $\bm{-}$ & $\bm{-}$ & $\bm{-}$ & $\bm{-}$ \\
\hline\hline\hline
\end{tabular*}
\\\vspace{1mm}\parbox{8.3cm}{Note: The second column lists the values of parameter \texttt{slevel}. \ding{51} and \ding{55} in the $(j+2)$-th column indicate whether the analysis produces Alarm $j$ (\ding{51}) or not (\ding{55})$\bm{-}$ marks a failed analysis.
}
}
\end{center}

{\parf} refines the distribution of $\Pbase$ based on its latticed sample spaces $(S, \sqsubseteq, \join, \meet, \top, \bot)$, leveraging all the collected intermediate results. Table 4 shows an example of such refinement for \texttt{slevel}, which is a crucial parameter for controlling the capacity of separate (unmerged) states during the static analysis. The individual analyses as exemplified in Table 4 are produced in parallel within a single iteration. Our artifact then constructs a matrix $\bm{R} \in \{\text{\ding{51}},\text{\ding{55}}\}^{m\times n}$, to represent the intermediate results (excluding failed analyses), where $m$ is the number of successfully completed analyses and $n$ is the cardinality of the universal set of reported alarms. We also use an $m$-dimensional vector $\bm{V}$ to denote the parameter values used in each analysis ($\bm{V}_i$ signifies the parameter value for the $i$-th analysis). Next, {\parf} performs Algorithm 1 to refine the distribution of $\Pbase$.

\begin{center}
\vspace{2mm}
\hrule height 1.2pt
\vspace{1mm}
{\begin{flushleft}\footnotesize{\bf Algorithm 1.} Refining the Distribution of $\Pbase$\end{flushleft}}
\vspace{1mm}
\hrule height 0.4pt
\vspace{1mm}

\hfil\begin{minipage}{.98\columnwidth}
\footnotesize{
    \SetKwInput{Input}{Input}\SetKwInOut{Output}{Output}\SetNoFillComment\SetNoFillComment
    \Input{$\bm{R}$: $m\times n$ intermediate-result matrix; $\bm{V}$: $m$-dimensional parameter value vector; $\Pbase$: original distribution.}
    \Output{$\Pbase^{'}$: refined distribution.}
    $\Pbase^{'} \leftarrow \Pbase$\,;
    
    \For(\tcp*[f]{$\mathtt{iterate\;\,over\;\,columns\;\,of}\;\,\bm{R}\;\,\mathtt{(alarms)}$}){$j \leftarrow 1$ \KwTo $n$}{
        $tmp \leftarrow \top$\,;
        
        \For(\tcp*[f]{$\mathtt{scan\;\,rows\;\,of}\;\,\bm{R}\;\,\mathtt{(analyses)}$}){$i \leftarrow 1$ \KwTo $m$}{
            \lIf{$\bm{R}_{ij} = \text{\ding{55}}$}{
                $tmp \leftarrow tmp \sqcap \bm{V}_i$ 
            }
        }
        \lIf{$tmp \neq \top$}{
            $\Pbase^{'} \leftarrow \Pbase^{'}\sqcup tmp$ 
        }
    }
    \Return $\Pbase^{'}$\,;
}
\end{minipage}
\vspace{1mm}
\hrule height 1.2pt
\vspace{2mm}
\end{center}

Algorithm 1 employs two nested loops. For the $j$-th column of $\bm{R}$ (w.r.t.\ Alarm $j$), the inner loop computes the greatest lower bound (for the lowest precision) of all sampled parameters which can eliminate (false) Alarm $j$. The outer loop casts the least upper bound for eliminating {all} such false alarms with the lowest precision.
Consider the example in Table 4, $\Pbase$ is refined as
\begin{align*}
    \Pbase^{'} \eeq \Pbase \,&\join\, (\top\meet 58\meet 103\meet104\meet1000\meet9) \\
            &\join\, (\top\meet 58\meet 103\meet104\meet1000) \\
            &\join\, (\top\meet104\meet1000), \\
            \eeq \Pbase &\join\, 9 \join 58 \join 104~.
\end{align*}%
It follows that $\Pbase^{'}$ is the {least precise} parameter setting (w.r.t.\ \verb|slevel|) that can eliminate {all} newly discovered false alarms in the current iteration.

For refining the distribution of $\Pdelta$, {\parf} uses the so-called completion rate $\eta_c$, i.e., the ratio of successfully completed analyses to all the $num_\text{sample}$ analyses. $\Pdelta$ is then refined via the scaling factor $\eta = 2\eta_c+\frac{1}{num_\text{sample}}$ as per Table 5 ($\eta > 1$ for $\eta_c \geq 0.5$). A {larger} value of $\eta$ indicates that more analyses have been completed within the allocated time budget, suggesting that a {more extensive exploration} of the parameter space (by scaling up $\Pdelta$) is possible, and vice versa.

\tabcolsep 4pt
\renewcommand\arraystretch{1.3}
\noindent
\begin{center}
\vspace{-1mm}
{\footnotesize{\bf Table 5.} Refining the Distribution of $\Pdelta$}\\
\vspace{2mm}
\footnotesize{
\begin{tabular*}{\linewidth}{ccp{0.4\linewidth}}\hline\hline\hline
Type & Original $\Pdelta$ & Refined $\Pdelta$ \\\hline
    Integer & $\text{Poisson}(\lambda)$ & $\text{Poisson}(\lambda\times \eta)$ \\
    Boolean & $\text{Bernoulli}(q)$ & $\text{Bernoulli}(1-(1-q)^{\eta})$ \\
    Set-of-Strings & $\text{B}(q_1)\times\ldots\times\text{B}(q_c)$ & $\text{B}(1-(1-q_1)^{\eta})\times\ldots\times\text{B}(1-(1-q_c)^{\eta})$ \\
\hline\hline\hline
\end{tabular*}
\\\vspace{1mm}\parbox{8.3cm}{Note: $\text{B}(q)$ is shorthand for $\text{Bernoulli}(q)$.}
}
\end{center}


\section{Empirical Evaluation}\label{sec:04-evaluation}
In this section, we evaluate the {\parf} artifact
\jz{3}{\url{https://hub.docker.com/repository/docker/parfdocker/parf-jcst/general}, May 2025.} to answer the following research questions:

\textit{RQ1 (Consistency).} Can the artifact reproduce experimental results as reported in~\cite{ase24-parf} (given the inherent randomness of {\parf} due to the sampling module)?

\textit{RQ2 (Verification Capability).} \revision{Can \parf improve \framac in verification competitions?}

\textit{RQ3 (Dominancy).} Which are the dominant (i.e., most influential) parameters in {\eva}?

\textit{RQ4 (Interpretability).} How does {\parf} help eliminate false alarms or even certify the absence of RTEs?

\setcounter{table}{5}
\begin{figure*}[t]
\begin{minipage}[b]{\textwidth}
\vspace{-2mm}
\begin{table}[H]
\centering
\caption{\label{tab:RQ1-reproduction-results}Experimental Results in Terms of RQ1 (Consistency)}
\centering\scriptsize
\renewcommand\arraystretch{1.0}
\begin{tabular}{lrrrrrrrrr}
\toprule
\multicolumn{4}{c}{\footnotesize{\textbf{OSCS Benchmark Details}}}
& \multicolumn{4}{c}{\footnotesize{\textbf{\#Alarms of Baselines}}}
& \multicolumn{2}{c}{\footnotesize{\textbf{\#Alarms of {\parf}}}} \\
\cmidrule(lr){1-4}\cmidrule(lr){5-8}\cmidrule(lr){9-10}
Benchmark name & LOC & \#Statements & \verb|-eva-precision| & \precision & \default & \expert & \official & \parfopt & \parfavg \\
\cmidrule(lr){1-4}\cmidrule(lr){5-8}\cmidrule(lr){9-10}
2048 & 440 & 329 & 6 & 13 & 7 & 5 & 7 & \best{4} & \gray{4.33} \\
chrony & 37177 & 41 & 11 & 9 & 9 & \similar{7} & 8 & \similar{7} & \gray{7.00} \\
debie1 & 8972 & 3243 & 2 & 33 & 33 & 3 & \best{1} & 2 & \gray{3.33} \\
genann & 1183 & 1042 & 10 & 236 & 236 & \similar{69} & 77 & \similar{69} & \gray{69.00} \\
gzip124 & 8166 & 4835 & 0 & 885 & 884 & 885 & 866 & \best{807} & \gray{836.00} \\
hiredis & 7459 & 87 & 11 & 9 & 9 & \similar{0} & 9 & \similar{0} & \gray{0.00} \\
icpc & 1302 & 424 & 11 & 9 & 9 & \similar{1} & \similar{1} & \similar{1} & \gray{1.00} \\
jsmn-ex1 & 1016 & 1219 & 11 & 58 & 58 & \similar{1} & \similar{1} & \similar{1} & \gray{1.00} \\
jsmn-ex2 & 1016 & 311 & 11 & 68 & 68 & \similar{1} & \similar{1} & \similar{1} & \gray{1.00} \\
kgflags-ex1 & 1455 & 474 & 11 & 11 & 11 & \similar{0} & 11 & \similar{0} & \gray{0.00} \\
kgflags-ex2 & 1455 & 736 & 10 & 33 & 33 & \similar{19} & 33 & \similar{19} & \gray{19.00} \\
khash & 1016 & 206 & 11 & 14 & 14 & \similar{2} & 14 & \similar{2} & \gray{2.00} \\
kilo & 1276 & 1078 & 2 & 523 & 523 & 445 & 688 & \best{419} & \gray{421.67} \\
libspng & 4455 & 2377 & 7 & 186 & 186 & \similar{122} & \similar{122} & 126 & \gray{145.33} \\
line-following-robot & 6739 & 857 & 10 & \similar{1} & \similar{1} & \similar{1} & \similar{1} & \similar{1} & \gray{1.00} \\
microstrain & 51007 & 3216 & 6 & 1177 & 1177 & 616 & 646 & \best{601} & \gray{606.00} \\
mini-gmp & 11706 & 628 & 6 & 83 & 83 & 71 & 83 & \best{65} & \gray{68.67} \\
miniz-ex1 & 10844 & 3659 & 1 & 2291 & 2291 & 1832 & 2291 & \best{1 }& \gray{763.67} \\
miniz-ex2 & 10844 & 5589 & 1 & 2748 & 2742 & \similar{2220} & 2742 & \similar{2219} & \gray{2475.33} \\
miniz-ex3 & 10844 & 3747 & 1 & 585 & 577 & 552 & 577 & \best{432} & \gray{510.67} \\
miniz-ex4 & 10844 & 1246 & 4 & 264 & 258 & 217 & 258 & \best{188} & \gray{206.67} \\
miniz-ex5 & 10844 & 3430 & 2 & 431 & 425 & \best{371} & 425 & 385 & \gray{389.00} \\
miniz-ex6 & 10844 & 2073 & 2 & 220 & 220 & 190 & 220 & \best{175} & \gray{183.33} \\
monocypher & 25263 & 4126 & 2 & 606 & 606 & \similar{564} & \similar{568} & 572 & \gray{577.67} \\
papabench & 12254 & 36 & 11 & \similar{1} & \similar{1} & \similar{1} & \similar{1} & \similar{1} & \gray{1.00} \\
qlz-ex1 & 1168 & 229 & 11 & 68 & 68 & \similar{11} & 68 & \similar{11} & \gray{21.33} \\
qlz-ex2 & 1168 & 75 & 11 & \similar{8} & \similar{8} & \similar{8} & \similar{8} & \similar{8} & \gray{8.00} \\
qlz-ex3 & 1168 & 294 & 8 & 94 & 94 & \similar{82} & 94 & \similar{82} & \gray{82.00} \\
qlz-ex4 & 1168 & 164 & 11 & 17 & 17 & \similar{13} & 17 & \similar{13} & \gray{13.00} \\
safestringlib & 29271 & 13029 & 7 & 855 & 855 & \best{256} & 300 & 263 & \gray{268.33} \\
semver & 1532 & 728 & 9 & 29 & 29 & \similar{22} & 25 & \similar{22} & \gray{23.00} \\
solitaire & 338 & 396 & 11 & 216 & 216 & \similar{18} & 213 & \similar{18} & \gray{18.00} \\
stmr & 781 & 500 & 6 & 63 & 63 & \similar{58} & 59 & \similar{58} & \gray{58.00} \\
tsvc & 5610 & 5478 & 4 & 413 & 413 & \similar{355} & 379 & \similar{354} & \gray{356.00} \\
tutorials & 325 & 89 & 11 & 5 & 5 & 1 & 5 & \best{0} & \gray{0.00} \\
tweetnacl-usable & 1204 & 659 & 11 & 126 & 126 & \similar{25} & 30 & \similar{25} & \gray{25.00} \\
x509-parser & 9457 & 3112 & 3 & 208 & 208 & 198 & 198 & \best{181} & \gray{185.33} \\
\midrule
\multicolumn{4}{l}{\footnotesize{Overall (\similar{tied-best}+\best{exclusively best})}} & \footnotesize{3/37} & \footnotesize{3/37} & \footnotesize{24/37} & \footnotesize{9/37} & \footnotesize{33/37 (89.2\%)} & \gray{--}\\
\multicolumn{4}{l}{\footnotesize{Overall (\best{exclusively best})}} & \footnotesize{0/37} & \footnotesize{0/37} & \footnotesize{2/37} & \footnotesize{1/37} & \footnotesize{11/37 (29.7\%)} & \gray{--} \\
\bottomrule
\end{tabular}
\end{table}
\end{minipage}
\end{figure*}

\subsection{Experimental Setup}

\vspace*{-2mm}
\paragraph*{Benchmarks.} We evaluate {\parf} over two benchmark suites:

1) The first suite is \framac Open Source Case Study (OSCS) Benchmarks\jz{4}{\url{https://git.frama-c.com/pub/open-source-case-studies}, May 2025.} (as per~\cite{ase24-parf}), comprising 37 real-world C code bases, such as the ``X509'' parser project (a {\framac}-verified static analyzer)~\cite{x509-parser} and ``chrony'' (a versatile implementation of the Network Time Protocol). The benchmark details are provided in \cref{tab:RQ1-reproduction-results}.

2) The second suite is collected from the verification tasks of SV-COMP 2024~\cite{svcomp2024}, where \framac participated in the NoOverflows category with a specific version called \framacsv~\cite{frama-c-sv}.


\vspace*{-2mm}
\paragraph*{Baselines.} We compare \parf against four parameter-tuning mechanisms: \precision, \default, \expert, and \official. The former two adopt the lowest-precision and default abstraction strategies of {\eva}, respectively. \expert dynamically adjusts the precision of abstraction strategies by sequentially increasing the \verb|-eva-precision| meta-option from 0 to 11 until the given time budget is exhausted or the highest precision level is reached. The \official mechanism uses the tailored strategies provided by {\eva} for the OSCS benchmarks, which can be regarded as \enquote{high-quality} configurations.

\vspace*{-2mm}
\paragraph*{Configurations.} All experiments are performed on a system equipped with two AMD EPYC 7542 32-core Processors and 128GB RAM running Ubuntu 22.04.5 LTS. To attain consistency, we adopt the same hyper-parameters as in~\cite{ase24-parf} (with $num_\text{sample} = 4$ and $time_\text{budget} = 1$ hour for each benchmark).

\begin{figure*}[t]
\begin{minipage}[b]{\textwidth}
    \begin{table}[H]
      \centering
      \caption{\label{tab:08-verification-capability}\revision{SV-COMP verification results in terms of RQ2 (Verification Capability)}}
        \centering\footnotesize
        \resizebox{\textwidth}{!}{ 
        \begin{tabular}{lcccccccc}
            \toprule
             & \multicolumn{7}{c}{\footnotesize{\textbf{Verification Result}}} & \\
            \cmidrule(lr){2-8}
            \textbf{Setting} & \multicolumn{2}{c}{Correct} & \multicolumn{2}{c}{Incorrect} & \multicolumn{3}{c}{Invalid} & \textbf{Score}  \\
            \cmidrule(lr){2-3}\cmidrule(lr){4-5}\cmidrule(lr){6-8}
            & True (\green{$+2$}) & False (\green{$+1$}) & True (\maroon{$-32$}) & False (\maroon{$-16$}) & Unknown (\gray{$0$}) & Failure (\gray{$0$}) & Error (\gray{$0$}) & \\
        \midrule
        ${\framacsv}_{\text{precision11}}$ & 1057 & 12 & 35 & 0 & 564 & 104 & 48 & 1006 \\
        ${\framacsv}_{\parf}$ & \textbf{1096} & 12 & 35 & 0 & 629 & \textbf{0} & 48 & \textbf{1084} \\
        \bottomrule
        \end{tabular}
        }
    \end{table}
\end{minipage}
\end{figure*}

\subsection{RQ1: Consistency}\label{subsec:RQ1}

\cref{tab:RQ1-reproduction-results} reports the analysis results in terms of the number of emitted alarms. For {\parf}, due to its inherent randomness, we repeat each experiment three times and report both the best result (\parfopt) and the averaged result (\parfavg). Since the former is also adopted in~\cite{ase24-parf}, we primarily compare {\parfopt} against the four baselines. We mark results with the exclusively fewest alarms (with difference $> 1\%$) as \best{exclusively best} and results with the same least number of alarms (modulo a difference of $\leq 1\%$) as \similar{tied-best}.

Overall, {\parf} achieves the least number of alarms on 33/37 (89.2\%) benchmarks with exclusively best results on 11/37 (29.7\%) cases, significantly outperforming its four competitors.
{These results are consistent with those obtained in~\textnormal{\cite{ase24-parf}}} (best: 34/37; exclusively best: 12/37). The minor differences stem primarily from the inherent randomness of {\parf} and changes in hardware configurations. We observe a special case for ``miniz-ex1'', where \#alarms reduces from 1828 as in~\cite{ase24-parf} to 1. This correlates with the fact that {\parf} finds an abstraction strategy that triggers a drastic decrease in {\framac}'s analysis coverage~\cite{buhlerEvaFramaC27}.
Moreover, as is observed in~\cite{ase24-parf}, {\parf} is particularly suitable for analyzing complex, large-scale real-world programs (i.e., benchmarks featuring low levels of \verb|-eva-precision|).


\subsection{\revision{RQ2: Verification Capability}}

Static analyzers, such as \framac, can be applied in verification scenarios~\cite{svcomp2024}. \cref{tab:08-verification-capability} shows that {\parf} can improve the performance of {\framac} in SV-COMP. \append{The detailed scoring schema is presented in~\cref{tab:08-verification-capability} of Appendix~\cref{appx-B:scoring-schema}, as per~\cite[Section 2]{svcomp2024}.} Since the analysis resource for each verification task is limited to 15 minutes of CPU time, ${\framacsv}_{\text{precision11}}$ strategy uses a fixed highest \texttt{-eva-precision 11} parameter for analysis (as per~\cite{frama-c-sv}). We set the hyper-parameters $time_\text{budget}$, $num_\text{process}$, and $num_\text{sample}$ of ${\framacsv}_{\parf}$ to 7.5 minutes, 2, 4.


\revision{The experimental results demonstrates {\parf}'s methodological robustness in enhancing verification capacity of {\framac}. Specifically, {\parf} eliminates all 104 analysis failures (due to the timeout) and verifies 39 more tasks, thereby improving the total score from 1006 to 1084.}
~Fig.~2 illustrates that {\parf} adaptively identifies 42 high-accuracy analysis results among the 104 failure cases, thus successfully verifying them. Furthermore, among all the 1057 true correct cases verified by ${\framacsv}_{\text{precision11}}$, {\parf} misses only 3.

\vspace{-1mm}
\begin{center}
\includegraphics[width=\linewidth]{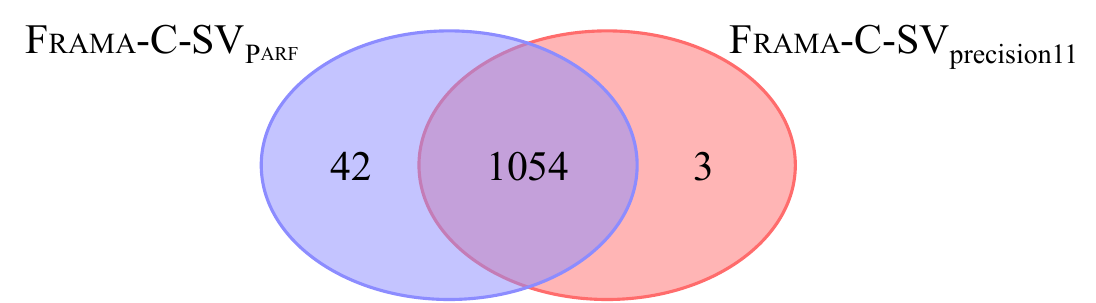}\\
\parbox[c]{8.3cm}{\footnotesize{Fig.~2.} \revision{Venn-diagram depicting the sets of true correct verification tasks by ${\framacsv}_{\parf}$ and ${\framacsv}_{\text{precision11}}$.}}
\end{center}

\subsection{RQ3: Dominancy}

We show via a post-hoc analysis that {\parf} supports the {identification of the most influential parameters} dominating the performance of {\eva}. To this end, we conduct 13 pairs (each for a single parameter) of controlled experiments for each OSCS benchmark\jz{6}{Trivial benchmarks where all parameter-tuning mechanisms yield identical performance (e.g., ``papabench'') are excluded.}. For instance, Fig.~3 depicts the results of 13 pairs of analyses for the ``2048'' benchmark. The analyses using the {\parfopt} configuration (reporting four alarms) and {\precision} configuration (reporting 13 alarms) signify a high-precision upper bound and a low-precision lower bound, respectively. For each parameter, we devise two types of controlled experiments: 1) \selected: The parameter is {selected} and retained from the {\parfopt} configuration, while the other 12 parameters are taken from the {\precision} configuration. This controlled experiment assesses the impact of the parameter w.r.t.\ the lower-bound baseline;

\noindent2) \excluded: The parameter is {excluded} from the {\parfopt} configuration and replaced with its counterpart from {\precision}, while the remaining 12 parameters are retained from the \parfopt configuration. This controlled experiment evaluates the parameter's influence w.r.t.\ the upper-bound baseline.

\vspace{-1mm}
\begin{center}
\includegraphics[width=\linewidth]{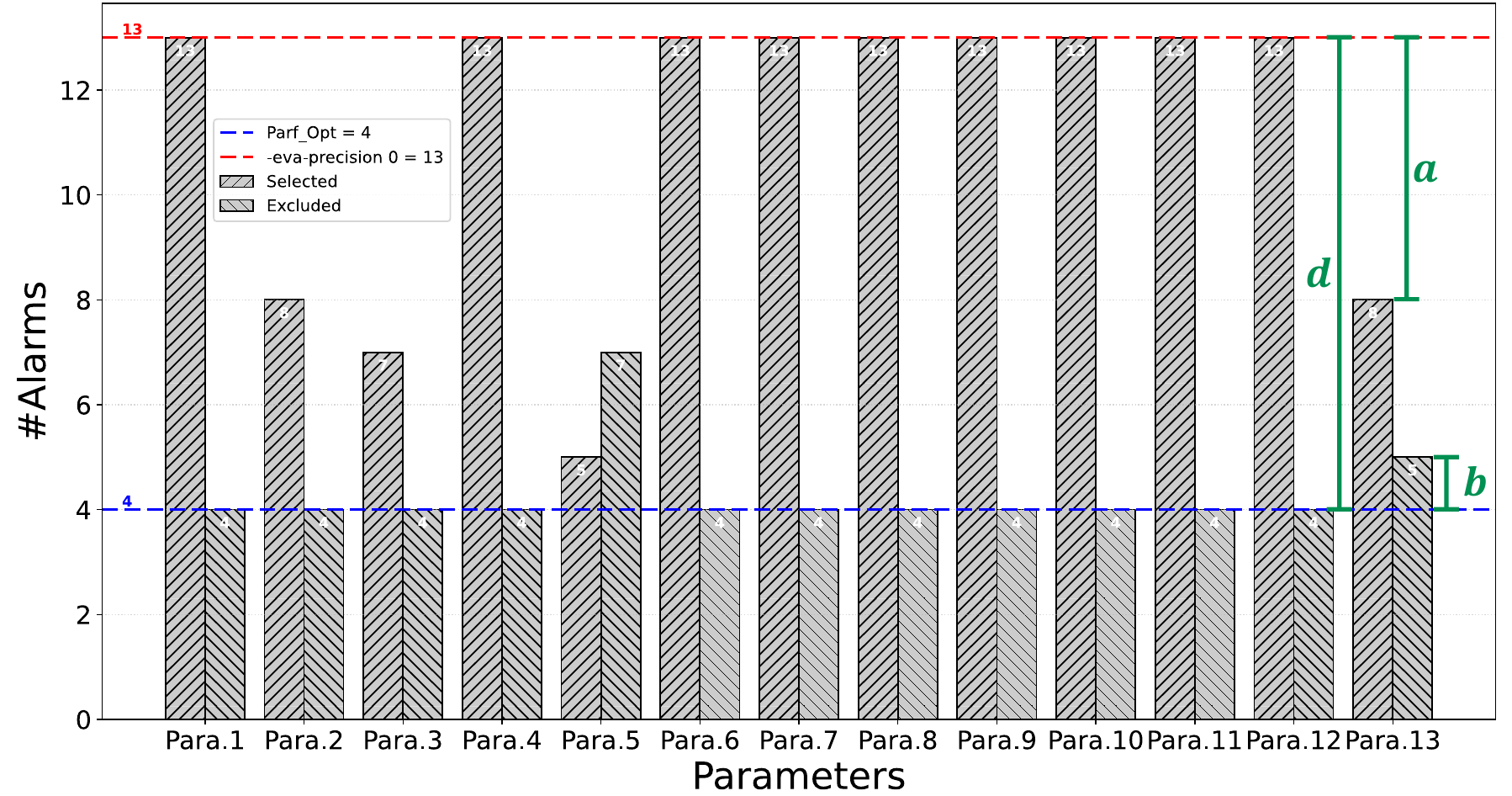}\\
\parbox[c]{8.3cm}{\footnotesize{Fig. 3.}  \#Alarms reported by analyses on the ``2048'' benchmark upon tuning individual parameters based on the abstraction strategies produced by \precision and \parfopt. Para.$n$ refers to the $n$-th parameter listed in Table 1.}
\end{center}

We then devise for each parameter a {scoring function} $s$ to quantitatively characterize its influence:
\begin{align*}\label{eq:composite-dist}
    s \ \ddefeq\ \frac{0.5 \cdot a \,+\, 0.5 \cdot b}{d}~,
\end{align*}%
where $a$, $b$, and $d$ capture the difference in \#alarms respectively between three cases: 1) the \precision baseline and the \selected experiment, 2) the \excluded experiment and the \parfopt baseline, and 3) the \precision baseline and the \parfopt baseline. For instance, for Para.$13$ in Fig.~3, we have $a = 13-8 = 5$, $b = 5-4 = 1$, and $d = 13-4 = 9$. 

\append{\cref{tab:09-parameters-impact} of Appendix~\ref{appx-C:RQ3-dominancy-details}} collects the influence score for all parameters across the OSCS benchmarks. For each benchmark, we mark the parameter with the highest score as the \dominant{dominant parameter}.
It follows that, overall, \verb|slevel| (Para.5) is the most influential parameter in {\eva} (with an averaged score of 0.490) and \verb|domains| (Para.13) is the second most influential parameter (with an averaged score of 0.258). This observation conforms to the crucial roles of \verb|slevel| and \verb|domains| in static analysis: The former restricts the number of abstract states at each control point and the latter determines the types of abstract representations used.

Nonetheless, the dominant parameter can vary for different target programs, e.g., the dominant parameters for ``debie1'' and ``miniz-ex6'' are \verb|auto-loop-unroll| (Para.2) and \verb|min-loop-unroll| (Para.1), respectively. This suggests that many false alarms emitted for these benchmarks can be eliminated through loop unrolling. Notably, ``miniz-ex6'' contains multiple nested loops that require extensive iterations to be fully unwound.

An unexpected observation is that certain parameters exhibit {negative} influence scores on a few benchmarks, such as Para.8 on ``jsmn-ex2'' and most parameters on ``qlz-ex3''. These negative scores arise when the \selected experiments produce more alarms than the \precision baseline, or when the \excluded experiments result in fewer alarms than the \parfopt baseline \append{(cf.~\cref{fig:05-dominant-parameters} of Appendix~\ref{appx-C:RQ3-dominancy-details})}. \revision{This phenomenon suggests that {\eva} is {not strictly monotonic} in certain cases. Nevertheless, our refinement mechanism equips {\parf} with the potential to handle these corner cases effectively. One such case is examined by the {\framac} community~\jz{7}{\url{https://stackoverflow.com/q/79497136/15322410}, May 2025} as well as discussed in Subsection~\ref{subsec:case-study}.}

\subsection{RQ4: Interpretability}\label{subsec:case-study}

We show how {\parf} helps eliminate false alarms or even certify the absence of RTEs through a case study. Fig.~4 gives a simplified version of the ``tutorials'' benchmark -- a toy program used to calculate differences between the ID of each parent process and its children.

\begin{center}
\lstset{
    language=C,                   
    basicstyle=\ttfamily\scriptsize,
    keywordstyle=\color{blue},    
    commentstyle=\color{gray}, 
    flexiblecolumns=true,
    escapechar=!,
    numbers=left,                  
    numberstyle=\tiny\color{gray}, 
    stepnumber=1,                  
    numbersep=5pt,                 
    frame=single,                  
    breaklines=true,               
    captionpos=b,                  
    xleftmargin=0pt,             
    xrightmargin=1pt              
}
\begin{lstlisting}[language=C,
    caption={Simplified version of benchmark ``tutorials''.},
    label={fig:case-study},
    title={Fig.4. A simplified version of the benchmark ``tutorials''.},
    % emph={@@assert p[p_id].child[i] < MAX_PROCESS_NUM;@@}, 
    % emphstyle=\color{red}\bfseries
]
#define MAX_CHILD_LEN 10
#define MAX_BUF_SIZE 100
#define MAX_PROCESS_NUM 50

struct process {
  uint8_t pid;
  uint8_t child_len;
  uint8_t child[MAX_CHILD_LEN];
};
struct process p[MAX_PROCESS_NUM];

// init returns -1 if initializing p[p_id] fails and 0 otherwise
int init(uint8_t *buf, uint16_t *offset, uint8_t p_id){
  ... // the concrete implementation body is abstracted away
};

int main(){
  uint8_t buf[MAX_BUF_SIZE], p_nb;
  uint16_t offset = 0;
  random_init((char*)buf, MAX_BUF_SIZE);

  // initialize the global array p of process structures
  for(p_nb = 0; p_nb < MAX_PROCESS_NUM; p_nb++){
    int r = init(buf, &offset, p_nb);
    if(r) break;
  }

  // print pid diff. btw. each valid process and its children
  for(uint8_t p_id = 0; p_id < p_nb; p_id++){
    for(uint8_t i = 0, c_id; i < p[p_id].child_len; i++)
      c_id = p[p_id].child[i];
      !\maroon{//@ assert c\_id < MAX\_PROCESS\_NUM;}!
      printf(" %i",p[p_id].pid - p[c_id].pid);
  }
  return 0;
};
\end{lstlisting}
\end{center}

\cref{tab:RQ1-reproduction-results} shows that {\parf} suffices to eliminate {all} alarms, yet \expert reports 1 false alarm. This alarm corresponds to the assertion \texttt{c\_id < MAX\_PROCESS\_NUM} in Line 32, signifying a potential out-of-bound RTE. A typical way to eliminate this false alarm is by maintaining a sufficiently large number of abstract states at this control point (loop condition in Line 30) by setting a high \verb|slevel| to prevent an over-approximation of the value of \texttt{c\_id}. This trick, unfortunately, does not work for this specific program (\expert sets \verb|slevel| to 5000, as is similar to \parf). The reason why {\parf} can eliminate the false alarm lies in its configuration of \verb|partition-history|: \expert sets it to 2, yet \parf sets it to 0. When \verb|partition-history| is set to $n\geq 1$, it delays the application of join operation on abstract domains, leading to an {exponential} increase (in $n$) of the number of abstract states required to avoid over-approximations at control points. Consequently, \expert using both high-precision \verb|slevel| and \verb|partition-history| fails to eliminate the false alarm in question, whilst \parf succeeds by pairing a high-precision \verb|slevel| with a low-precision \verb|partition-history|.

The effectiveness of \parf roots in its ability to maintain low-precision distributions for {disturbing} parameters (those with negative contributions to eliminating false alarms for specific programs, e.g., \verb|partition-history| for ``tutorials'') while achieving high-precision distributions for {dominant} parameters (e.g., \verb|slevel| for ``tutorials'') during the refinement procedure. This ingenuity can be attributed to two key factors: 1) Unlike \expert, which groups and binds all parameters into several fixed configuration packs, \parf models each parameter as an independent random variable; 2) The \enquote{meet-and-join} refinement strategy (described in Algorithm 1) restrains the growth of $\Pbase$ for disturbing parameters while increasing $\Pbase$ for dominant parameters. In a nutshell, despite its assumption on monotonic analyzers (cf.\ \cref{sec:02-method}), \parf exhibits strong potential to improve the performance of static analyzers that lack strict monotonicity.
\section{\revision{Limitations and Future Work}}\label{sec:05-limitation-future-work}

\revision{We pinpoint several scenarios for which {\parf} is inadequate and provide potential solutions thereof.}

\revision{First, {\parf} models different parameters of a static analyzer as independent random variables. However, the interactions between parameters can potentially lead to complex parameter dependencies. For instance, 
1) larger \texttt{partition-history} requires (exponentially) larger \texttt{slevel} to delay approximations for all conditional structures~\cite[Subsection 6.5.1]{buhlerEvaFramaC27}, and 2) the modification of \texttt{domains} can unpredictably interact with \texttt{slevel}~\cite[Subsection 6.7]{buhlerEvaFramaC27}. Taking into account the dependencies between parameters is expected to reduce the search space and thereby accelerate the parameter refining process. To this end, we need to extend {\parf} to admit the {representation of stochastic dependencies}, such as conditional random variable models.}

\revision{Second, parameter initialization in {\parf} relies on fixed heuristically-defined distributions, without leveraging historical experience (e.g., expert knowledge on configuring typical programs) or program-specific features (e.g., the syntactic or semantic characteristics of the source program) to optimize initial configurations. While neural networks or fine-tuned large language models could automate this process, their deployment requires balanced training data pairing programs with optimal parameters -- a dataset traditionally requiring expert curation. Notably, {\parf}'s automated configuration generation capability paves the way for constructing such datasets at scale, enabling data-driven initialization as promising future work.}

\revision{Third, while {\parf} enhances static analyzers' capacity, it cannot fully eliminate false positives due to the fundamental precision-soundness tradeoffs of static analysis. As shown in \cref{tab:RQ1-reproduction-results}, residual alarms require manual inspection. A promising direction involves integrating {\parf} with formal verification tools (e.g., proof assistants or SMT solvers) to classify alarm validity.}
\section{Related Work}\label{sec:06-related-work}


\paragraph*{Abstraction Strategy Refinement.} Beyer \textit{et al.}~\cite{ase08-dynamic-precision-of-PA} proposed CPA+, a framework that augmented the program verifier CPA~\cite{cav07-CPA} with deterministic abstraction-strategy tuning schemes based on intermediate analysis information (e.g., predicates and abstract states). CPA+ aims to enhance the scalability and efficiency of verification, such as predicate abstraction-based model checking, while \parf focuses on improving the accuracy of static analysis leveraging the alarm information. Zhang \textit{et al.}~\cite{oopsla24-selecting-abstraction} introduced BinGraph, a framework for learning abstraction selection in Bayesian program analysis. Yan \textit{et al.}~\cite{oopsla24-scaling-abstraction} proposed a framework that utilized graph neural networks to refine abstraction strategies for Datalog-based program analysis. These data-driven methods~\cite{oopsla24-selecting-abstraction,oopsla24-scaling-abstraction} require datasets for training a Bayesian/neural network, while \parf requires no pre-training effort.
The theoretical underpinnings of \parf are established in~\cite{ase24-parf}, we focus on the design, implementation, and application of \parf and make new contributions detailed in \cref{sec:01-intro}.

\paragraph*{\revision{Improving Static Analyzers.}}
\revision{Modern static analyzers employ diverse parameterization strategies to balance precision and performance.}
\revision{K\"astner \textit{et al.}~\cite{kaestnerAutomaticSoundStatic2023} summarized the four most important abstraction mechanisms in Astr{\'e}e and recommended prioritizing the accuracy of related abstract domains, which amounts to narrowing down the parameter space. However, these mechanisms need hand-written directives and thus are not fully automated. \textsc{Mopsa}~\cite{tacas24:mopsa} adopts a fixed sequence of increasingly precise configurations akin to \eva's \expert mechanism when participating SV-COMP 2024. \parf can be generalized to \textsc{Mopsa} by modeling its specific parameters, thus helping to decide the best configuration to analyze a given program. Saan \textit{et al.}~\cite{saanGoblintAutotuningThreadModular2023} implemented in \textsc{Goblint} a simple, heuristic autotuning method based on syntactical criteria, which can activate or deactivate abstraction techniques before analysis. However, this method only generates an initial analysis configuration once and does not dynamically adapt to refine the parameter configuration.}
\section{Conclusions}\label{sec:07-conclusion}

We have presented the {\parf} toolkit for adaptively tuning abstraction strategies of static program analyzers. It is -- to the best of our knowledge -- the first {fully automated} approach that supports {incremental} refinement of such strategies.
The effectiveness of {\parf} has been demonstrated through a case study and collections of standard benchmarks and SV-COMP 2024 tasks. Interesting future directions include extending {\parf} to cope with dependencies between parameters, neural network-based parameter initialization, and combining formal verification tools.


\paragraph*{\bf Conflict of Interest.}
The authors declare that they have no conflict of interest.

\vspace{5mm}

\newpage
\begin{appendices}
\renewcommand{\thesection}{A\arabic{section}}
\renewcommand{\thetable}{A\arabic{table}}

\section{Initial Probability Distributions}
\label{appx-A:initial-prob-dist}
The table 8 displays the initial distributions employed by the artifact. The choice of $\Pbase$ aligns with the configuration \verb|-eva-precision| 0, which serves as a low-precision starting point for incremental refinement. $\Pdelta$'s for Boolean-typed parameters are subject to Bernoulli distributions modeling fair coin flips. In contrast, the initialization of $\Pdelta$ for integer-typed parameters reflects the efficiency impact of different parameters: parameters with higher computational costs (indicated by smaller values in \verb|-eva-precision| 11) follow Poisson distributions with smaller expected values.

For simplicity, the Dirac-distributed $\Pbase$ is denoted by its unique non-zero support, e.g., $\Pr[\Pbase=10]=1$ for integer parameter \texttt{plevel}, $\Pr[\Pbase=\bFALSE]=1$ for Boolean parameter \texttt{split-return}, $\Pr[\Pbase=(\bTRUE,\bFALSE,\bFALSE,\bFALSE,\bFALSE)]=1$ for set-of-strings parameter \texttt{domains}.

\tabcolsep 6pt
\renewcommand\arraystretch{1.3}
\noindent
\begin{center}
\vspace{-1mm}
{\footnotesize{\bf Table A1.} Initial Probability Distributions of Parameters}\\
\vspace{2mm}
\footnotesize{
\begin{tabular*}{\linewidth}{lcc}\hline\hline\hline
        Parameter & $\Pbase$ & $\Pdelta$\\\hline
        \verb|min-loop-unroll| & 0 & $\text{Poisson}(0.4)$ \\
        \verb|auto-loop-unroll| & 0 & $\text{Poisson}(10)$ \\
        \verb|widening-delay| & 1 & $\text{Poisson}(0.5)$ \\
        \verb|partition-history| & 0 & $\text{Poisson}(0.4)$ \\
        \verb|slevel| & 0 & $\text{Poisson}(20)$ \\
        \verb|ilevel| & 8 & $\text{Poisson}(2)$ \\
        \verb|plevel| & 10 & $\text{Poisson}(10)$ \\
        \verb|subdivide-non-linear| & 0 & $\text{Poisson}(2.5)$ \\
        \verb|split-return| & F & $\text{Bernoulli}(0.5)$ \\
        \verb|remove-redundant-alarms| & F & $\text{Bernoulli}(0.5)$ \\
        \verb|octagon-through-calls| & F & $\text{Bernoulli}(0.5)$ \\
        \verb|equality-through-calls| & F & $\text{Bernoulli}(0.5)$ \\
        \verb|domains| & (T, F, F, F, F) & $\text{Bernoulli}(0.5)^5$ \\
\hline\hline\hline
\end{tabular*}
}
\end{center}

\section{Scoring Schema of RQ2}
\label{appx-B:scoring-schema}
Static analyzers, such as \framac, can be applied in verification scenarios. Each program in the SV-COMP NoOverflows category is either safe (i.e., correct) or contains an instance of integer overflow bug (i.e., incorrect). An analyzer successfully proves a task when reports zero alarms for a correct program (i.e., true correct), or a sure error for an incorrect program (i.e., false correct). Conversely, an analyzer is penalized if it falsely reports an error for a correct program (i.e., false incorrect) or fails to detect an error for an incorrect program (i.e., true incorrect). Additionally, analyses that generate uncertain alarms, exceed the time limit or encounter exceptions are classified as unknown, failure, and error, respectively. The detailed scoring schema is presented in~\cref{tab:08-verification-capability}, as per ~\cite[Section 2]{svcomp2024}.

\section{More Details on RQ3 (Dominancy)}
\newpage
\label{appx-C:RQ3-dominancy-details}

\setcounter{table}{1}
\begin{figure*}[t]
\begin{minipage}[b]{\textwidth}
    \begin{table}[H]
      \centering
        \caption{\label{tab:09-parameters-impact}Impact of Parameters in Terms of RQ3 (Dominancy)}
        \centering\footnotesize
        \scriptsize 
        \resizebox{\textwidth}{!}{ 
        \renewcommand\arraystretch{1.0}
        \begin{tabular}{lrrrrrrrrrrrrr}
            \toprule
            \textbf{Benchmark name} & \textbf{Para.1} & \textbf{Para.2} & \textbf{Para.3} & \textbf{Para.4} & \textbf{Para.5} & \textbf{Para.6} & \textbf{Para.7} & \textbf{Para.8} & \textbf{Para.9} & \textbf{Para.10} & \textbf{Para.11} & \textbf{Para.12} & \textbf{Para.13} \\
        \midrule
        2048           & 0.000 & 0.278 & 0.333 & 0.000 & \textbf{0.611} & 0.000 & 0.000 & 0.000 & 0.000 & 0.000 & 0.000 & 0.000 & 0.333 \\
        chrony         & 0.000 & 0.000 & 0.000 & 0.000 & 0.000 & 0.250 & 0.000 & 0.000 & 0.000 & 0.000 & 0.000 & 0.000 & \textbf{0.500} \\
        debie1         & 0.000 & \textbf{0.500} & 0.016 & 0.000 & 0.484 & 0.177 & 0.000 & 0.000 & 0.000 & 0.000 & 0.000 & 0.000 & 0.226 \\
        genann         & 0.287 & 0.009 & 0.000 & 0.000 & \textbf{0.665} & 0.000 & 0.000 & 0.009 & 0.006 & 0.000 & 0.012 & 0.000 & 0.024 \\
        gzip124        & 0.000 & 0.000 & 0.000 & 0.000 & 0.205 & -0.006 & 0.000 & 0.013 & 0.000 & 0.000 & 0.000 & 0.000 & \textbf{0.782} \\
        hiredis        & 0.000 & 0.000 & 0.000 & 0.222 & \textbf{0.722} & 0.000 & 0.000 & 0.000 & 0.000 & 0.000 & 0.000 & 0.000 & 0.167 \\
        icpc           & 0.000 & 0.000 & 0.000 & 0.000 & \textbf{0.500} & 0.000 & 0.000 & 0.250 & 0.000 & 0.000 & 0.250 & 0.000 & 0.250 \\
        jsmn-ex1       & 0.000 & 0.105 & 0.026 & 0.000 & \textbf{0.851} & 0.000 & 0.000 & -0.009 & 0.000 & 0.000 & 0.000 & 0.000 & 0.105 \\
        jsmn-ex2       & 0.000 & 0.000 & 0.000 & 0.000 & \textbf{0.993} & 0.000 & 0.000 & -0.030 & 0.000 & 0.000 & 0.000 & 0.000 & 0.045 \\
        kgflags-ex1    & 0.000 & 0.000 & 0.000 & 0.000 & \textbf{0.500} & 0.000 & 0.000 & 0.000 & 0.000 & 0.000 & 0.000 & 0.000 & 0.000 \\
        kgflags-ex2    & 0.000 & 0.000 & 0.000 & 0.000 & 0.107 & 0.000 & \textbf{0.786} & 0.000 & 0.500 & 0.000 & 0.000 & 0.000 & 0.000 \\
        khash          & 0.000 & 0.000 & 0.000 & 0.292 & \textbf{0.667} & 0.000 & 0.000 & 0.000 & 0.042 & 0.000 & 0.000 & 0.000 & 0.000 \\
        kilo           & 0.010 & 0.010 & 0.014 & 0.019 & 0.197 & 0.010 & 0.010 & 0.010 & 0.010 & 0.010 & 0.014 & 0.010 & \textbf{0.861} \\
        libspng        & 0.067 & 0.000 & 0.000 & 0.000 & \textbf{0.375} & 0.150 & 0.000 & 0.000 & 0.008 & 0.000 & 0.000 & 0.000 & 0.175 \\
        microstrain    & 0.000 & 0.457 & 0.000 & 0.000 & \textbf{0.503} & 0.000 & 0.000 & 0.000 & 0.040 & 0.000 & 0.000 & 0.000 & 0.000 \\
        mini-gmp       & 0.000 & 0.000 & 0.000 & 0.000 & 0.167 & 0.000 & 0.000 & 0.000 & 0.000 & 0.000 & 0.000 & 0.000 & \textbf{0.833} \\
        miniz-ex1      & 0.000 & 0.003 & 0.000 & 0.000 & \textbf{0.996} & 0.000 & 0.000 & 0.000 & 0.004 & 0.000 & 0.000 & 0.000 & 0.102 \\
        miniz-ex2      & 0.000 & 0.024 & 0.006 & 0.000 & 0.045 & 0.003 & 0.000 & 0.000 & 0.000 & 0.000 & 0.000 & 0.000 & \textbf{0.935} \\
        miniz-ex3      & 0.000 & 0.056 & 0.033 & 0.000 & \textbf{0.310} & 0.114 & 0.000 & 0.000 & 0.304 & 0.000 & 0.000 & 0.000 & 0.225 \\
        miniz-ex4      & 0.000 & 0.132 & 0.086 & 0.270 & \textbf{0.276} & 0.204 & 0.000 & 0.000 & 0.000 & 0.000 & 0.000 & 0.000 & 0.257 \\
        miniz-ex5      & 0.000 & 0.076 & 0.076 & 0.000 & 0.130 & 0.000 & 0.000 & 0.000 & 0.000 & 0.000 & 0.000 & 0.000 & \textbf{0.772} \\
        miniz-ex6      & \textbf{0.489} & 0.011 & 0.122 & 0.000 & 0.189 & 0.000 & 0.044 & 0.100 & 0.000 & 0.000 & 0.000 & 0.000 & 0.122 \\
        monocypher     & 0.074 & 0.441 & 0.000 & 0.000 & \textbf{0.485} & 0.000 & 0.000 & 0.059 & 0.000 & 0.000 & 0.000 & 0.000 & 0.000 \\
        qlz-ex1        & 0.044 & 0.000 & 0.000 & 0.044 & \textbf{1.000} & 0.000 & 0.000 & 0.000 & 0.000 & 0.000 & 0.000 & 0.000 & 0.035 \\
        qlz-ex3        & -0.167 & -0.167 & -0.083 & -0.167 & \textbf{0.542} & -0.167 & -0.167 & -0.167 & -0.167 & -0.167 & -0.167 & -0.167 & 0.458 \\
        qlz-ex4        & 0.000 & 0.000 & 0.000 & 0.000 & \textbf{1.000} & 0.000 & 0.000 & 0.000 & 0.000 & 0.000 & 0.000 & 0.000 & 0.000 \\
        safestringlib  & 0.000 & 0.267 & 0.001 & 0.000 & \textbf{0.655} & 0.004 & 0.000 & 0.000 & 0.000 & 0.000 & 0.000 & 0.000 & 0.132 \\
        semver         & 0.000 & 0.143 & 0.000 & 0.286 & \textbf{0.571} & 0.000 & 0.000 & 0.000 & 0.071 & 0.000 & 0.000 & 0.000 & 0.000 \\
        solitaire      & 0.482 & 0.000 & 0.000 & 0.000 & \textbf{0.505} & 0.000 & 0.000 & 0.000 & 0.000 & 0.000 & 0.000 & 0.000 & 0.106 \\
        stmr           & 0.000 & 0.000 & 0.000 & 0.000 & 0.100 & 0.000 & 0.000 & 0.000 & 0.000 & 0.000 & 0.000 & 0.000 & \textbf{0.800} \\
        tsvc           & 0.000 & 0.008 & 0.000 & 0.000 & \textbf{0.653} & 0.000 & 0.000 & -0.017 & 0.000 & 0.000 & 0.000 & 0.000 & 0.364 \\
        tutorials      & 0.000 & 0.200 & 0.000 & 0.000 & \textbf{0.700} & 0.000 & 0.000 & 0.100 & 0.100 & 0.000 & 0.000 & 0.000 & 0.000 \\
        tweetnacl-usable & 0.064 & 0.485 & 0.000 & 0.149 & \textbf{0.505} & 0.000 & 0.000 & -0.005 & 0.000 & 0.000 & 0.000 & 0.000 & 0.010 \\
        x509-parser    & 0.019 & 0.056 & 0.019 & 0.407 & \textbf{0.444} & 0.019 & 0.019 & 0.019 & 0.019 & 0.019 & 0.019 & 0.019 & 0.148 \\
        \midrule
        Average        & 0.040 & 0.091 & 0.019 & 0.045 & \textbf{0.490} & 0.022 & 0.020 & 0.010 & 0.028 & -0.004 & 0.004 & -0.004 & 0.258 \\
        \bottomrule
        \end{tabular}
        }
        \\\vspace{1mm}\parbox{17.5cm}{~~Para.$n$ refers to the $n$-th parameter listed in Table 1.}
    \end{table}
\end{minipage}
\end{figure*}

\newpage
\setcounter{figure}{3}
\begin{figure}[H]
\centering
\includegraphics[width=2.07\linewidth]{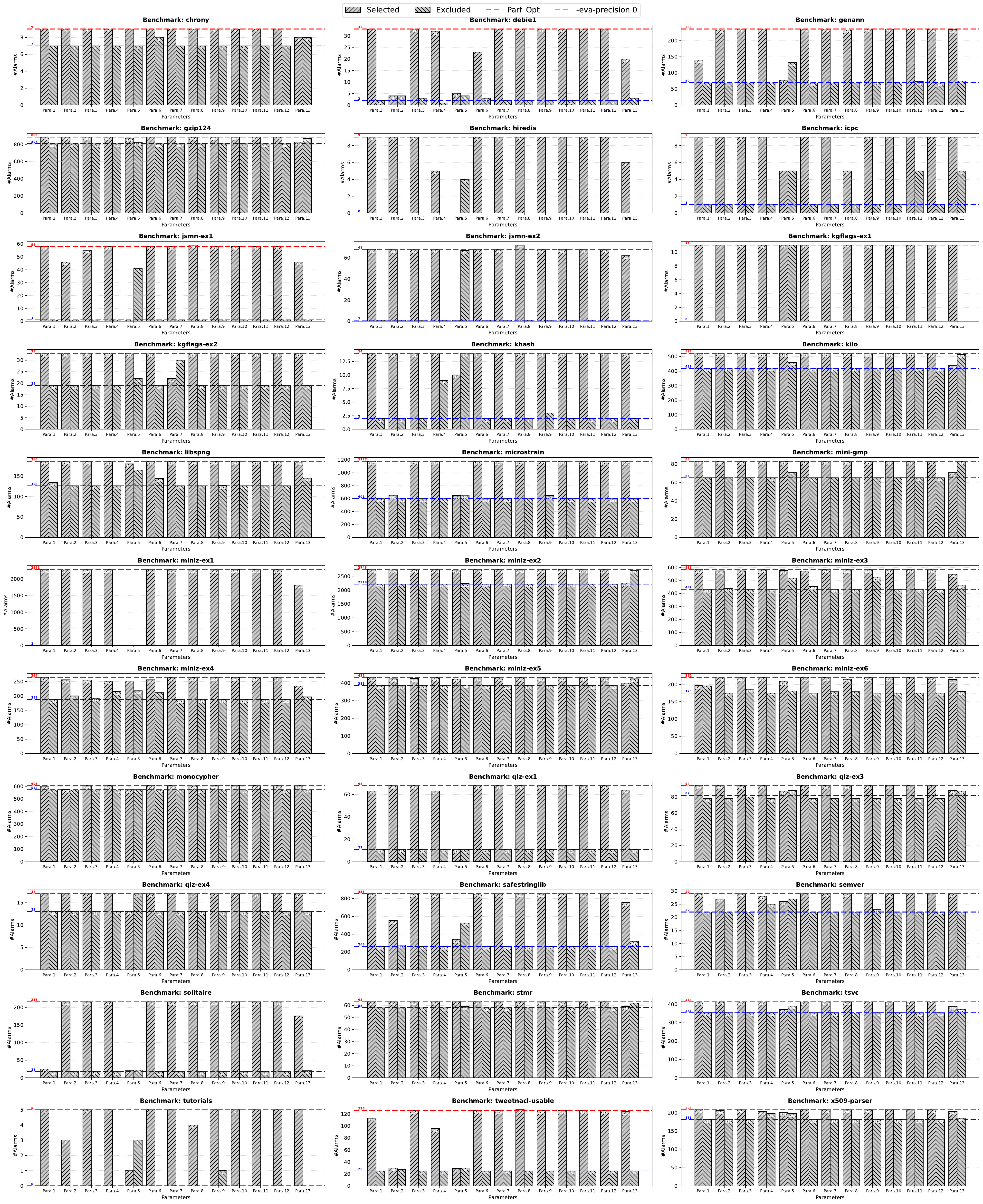}
\captionsetup{width=2\linewidth}
\caption{
\begin{minipage}{2.9\columnwidth}
\centering
\label{fig:05-dominant-parameters}{\footnotesize Fig. A1. Analysis results by selecting or excluding individual parameters on OSCS benchmarks.}
\end{minipage}
}
\end{figure}

\end{appendices}

\label{last-page}
\end{multicols}
\label{last-page}
\end{document}